\documentstyle[aps]{revtex}

\newcommand{\pcfr}{\mbox{$P_{cf}$}}
\newcommand{\exf}[2]{\mbox{$#1\!\times\! 10^{#2}$}}

\begin{document}
\draft
\title{Simulating temporal evolution of pressure 
in two-phase flow in porous media} 
\author{Eyvind Aker\thanks{Also at: Norwegian University of Science and Technology, N-7034 Trondheim, Norway}, Knut J\o rgen  M\aa l\o y}
\address{Department of Physics, University of Oslo, N-0316 Oslo, Norway}
\author{Alex Hansen\thanks{Also at: IKU Petroleum Research, N-7034 Trondheim, Norway}}
\address{Department of Physics, Norwegian University of Science and 
Technology, N-7034 Trondheim, Norway}
\date{\today}
\maketitle

\begin{abstract}
We have simulated the temporal evolution of pressure due to capillary
and viscous forces in two-phase drainage in porous media. We analyze
our result in light of macroscopic flow equations for two-phase
flow. We also investigate the effect of the trapped clusters on the
pressure evolution and on the effective permeability of the system. We
find that the capillary forces play an important role during the
displacements for both fast and slow injection rates and both when the
invading fluid is more or less viscous than the defending fluid. The
simulations are based on a network simulator modeling two-phase
drainage displacements on a two-dimensional lattice of tubes.
\end{abstract}


\section{Introduction}
Fluid flow in porous media such as sand, soil and fractured rock is an
important process in nature and has a huge number of practical
applications in engineering. Most often mentioned is oil
recovery and hydrology. Fluid flow in porous media has also been of
great interest in modern physics. In particular, the different
structures of the interface between the fluids in two-phase displacements have
been extensively studied. Despite this attention there are still many
open questions concerning fluid flow in porous media.

In this paper we report on simulations of the temporal
evolution of pressure during two-phase drainage in a model
porous medium, consisting of a two-dimensional lattice of
tubes. The network model has been developed to measure the time
dependence of different physical properties and to study the dynamics
of the fluid movements. Especially, we focus on the dynamics of the
temporal evolution of the pressure due to capillary and viscous
forces and the time dependence of the front between the two liquids.
The discussion is restricted to drainage displacement, i.e.\ the
process where a non-wetting fluid displaces a wetting fluid in a
porous medium.

During the last two decades an interplay between experimental
results and numerical simulations has improved the understanding of
the displacement process.  It has been shown that the different
structures observed when changing the physical parameters of the
fluids like viscosity contrast, wettability, interfacial tension and
displacement
rate~\cite{Maloy85,Chen-Wilk85,Len88,Cieplak88,Cieplak90,Cieplak91}
divide into three flow regimes.  These three major regimes are referred
to as viscous fingering~\cite{Maloy85,Chen-Wilk85} stable
displacement~\cite{Len88} and capillary fingering~\cite{Len85}. There
exist statistical models such as DLA~\cite{Witten81},
anti-DLA~\cite{Pater84} and invasion percolation~\cite{Wilk83} that
reproduce the basic domains in viscous fingering, stable displacement
and capillary fingering respectively.  However, these models do not
contain any physical time for the front evolution and they cannot
describe the crossover between the different flow regimes.  

Much effort has gone into making better models whose properties are
closer to that of real porous media. This has resulted in several
network simulators, modeling fluid flow on a lattice of pores and
throats~\cite{Chen-Wilk85,Len88,Kop-Lass85,%
Paya86-1,King87,Blunt90,Blunt91,Reeves96,Paya96,Oeren96,Blunt97}.
Most of the network models have been used to obtain new information
on the different flow regimes and to study the statistical
properties of the displacement structures. Some others, have been used
to calculate macroscopic properties like fluid saturations and
relative permeabilities and compare them with corresponding
experimental data. However, to our knowledge no network model has been
capable of simulating the dynamics of the fluid pressures during the
displacement process. This is the goal of the present work.

We have simulated the temporal evolution of the pressure in all the
three regimes of interest: viscous fingering, stable displacement and
capillary fingering. The injection rate in the displacements has been
systematically varied and we have analyzed the behavior of the
pressure in the crossover between the three regimes.  
Moreover, we discuss what effect trapped clusters have on the
evolution of the pressure in the system and we relate the data to
macroscopic flow equations.  We find that capillary forces play an
important role in two-phase flow at both high and low injection rates.

The paper is organized as follows. In Sec.\ \ref{sec:model} we present
the network model, in Sec.\ \ref{sec:result} we show the simulation
results and in Sec.\ \ref{sec:discuss} we present our 
conclusions and related the findings to macroscopic quantities.

\section{The Network Model}
\label{sec:model}
The network model has been presented in~\cite{Aker97}
and we will restrict ourself to a short sketch here.

\subsection{Geometry and boundary conditions}
The porous medium is represented by a square lattice tubes connected
at the nodes. At each node four tubes meet and there is no volume
assigned to the nodes: the tubes represent the volume of both pores
and throats.  The tubes are cylindrical with length $d$. Each tube is
assigned an average radius $r$ which is chosen at random in the
interval \mbox{$[\lambda_1 d,\lambda_2 d]$}, where $0\le\lambda_1 <
\lambda_2\le 1$.  The randomness of the radii represents the disorder
in the model.

The liquids flow from the bottom to the top of the lattice and we
implement periodic boundary conditions in the horizontal
direction. The pressure difference between the bottom row and the top
row defines the pressure across the lattice.  Gravity effects are
neglected, and consequently we consider horizontal flow in a
two-dimensional network of tubes.

\subsection{Fluid flow through the network}
Initially, the system is filled with a defending fluid with viscosity
$\mu_1$. The invading fluid with viscosity $\mu_2$ is injected along
the bottom row with a constant injection rate.  We model drainage,
i.e.\ the invading fluid is non-wetting and the defending fluid is
wetting.  Furthermore, we assume that the fluids are incompressible and
immiscible.  Consequently, the volume flux is conserved everywhere in the
lattice and a well defined interface develops between the two phases.

The capillary pressure $p_c$ due to the interface between the
non-wetting and wetting fluid inside a tube (a meniscus)
is given by the Young-Laplace law
\begin{equation}
p_c=\frac{2\gamma}{r}\cos\theta\ .
\label{eq:pc}
\end{equation}
Here $r$ is the radius of the tube, $\gamma$ is the interfacial tension
and $\theta$ denotes the wetting angle between the non-wetting and wetting
phases. $\theta$ is in the interval $(0,\pi/2)$ for drainage displacements.

We assume that the tubes are hour glass shaped where the effective
radii of the tubes follow a smooth function. Thus, the capillary
pressure becomes a function of the position of the meniscus in the
tube and we assume that the Young-Laplace law~(\ref{eq:pc}) takes the form
\begin{equation}
p_c=\frac{2\gamma }{r}\left[ 1-\cos (2\pi\hat{x})\right]\ .
\label{eq:pcvary}
\end{equation}
Here $\hat{x}$ is the dimensionless value of the meniscus' position in the tube
($0\le\hat{x}\le 1$), and $\theta=0$ (perfect
wetting).  From Eq.\ (\ref{eq:pcvary}) $p_c=0$ at the ends of the tube
while $p_c$ approaches its maximum of $4\gamma/r$ in the middle of the
tube.

The volume flux $q_{ij}$ through a tube from the $i$th to the $j$th
node in the lattice (Fig.~\ref{fig:tube}) is found from the Washburn
equation for capillary flow~\cite {Wash21}. As an approximation we
treat the tubes as if they were cylindrical and obtain
\begin{equation}
q_{ij}=-\frac{\pi r_{ij}^2 k_{ij}}{\mu_{eff}}\cdot \frac{1}{d}(\Delta p_{ij}-\tilde{p}_{c})\ .
\label{eq:tubeflow}
\end{equation}
Here $k_{ij}$ is the permeability of the tube given by
$k_{ij}=r_{ij}^2/8$ where $r_{ij}$ is the average radius of the tube.
$\Delta p_{ij}=p_j-p_i$ is the pressure difference between the $i$th
and $j$th node. A tube partially filled with both of the liquids is
allowed to contain either one or two menisci leading to four different
arrangements as shown in Fig.~\ref{fig:tubeconf}.  The effective
viscosity of those tubes, denoted as $\mu_{eff}$ in Eq.\
(\ref{eq:tubeflow}), becomes a sum of the amount of each fluid
multiplied by their respective viscosities. The total capillary
pressure, $\tilde{p}_c$ in Eq.\ (\ref{eq:tubeflow}), is the sum of the
capillary pressures of the menisci that are inside the tube. The
absolute value of the capillary pressure of each meniscus is given by
Eq.\ (\ref{eq:pcvary}), while its sign depends on whether the meniscus
is pointing upwards like in Fig.~\ref{fig:tubeconf}\,(a) or downwards
like in Fig.~\ref{fig:tubeconf}\,(b).  In the simple case where the
tube only contains one meniscus (Fig.~\ref{fig:tube}) $\mu_{eff}=\mu_2
\hat{x}_{ij}+ \mu_1(1-\hat{x}_{ij})$ and $\tilde{p}_c=p_c$.  For a
tube without menisci $\tilde{p}_c=0$, and Eq.\ (\ref{eq:tubeflow})
reduces to that describing Hagen-Poiseuille flow with
$\mu_{eff}=\mu_1\mbox{ or }\mu_2$.

\subsection{Determining the flow field}
There is no volume assigned to the nodes giving conservation of
volume flux at each node
\begin{equation}
\sum_j q_{ij}=0\ .
\label{eq:Kirch}
\end{equation}
The summation on $j$ runs over the nearest neighbor 
nodes to the $i$th node while $i$ runs over all nodes that do not 
belong to the top or bottom rows, that is, the internal nodes. 

Eqs.\ (\ref{eq:tubeflow}) and~(\ref{eq:Kirch}) constitute a set of
linear equations which are to be solved for the nodal pressures
$p_j$ with the constraint that the
pressures at the nodes belonging to the upper and lower rows are kept
fixed.  The set of equations is solved by using the Conjugate Gradient
method~\cite{CGM}.

We want to study the dynamics of the pressure fluctuations at constant
displacement rate. Therefore, we need to find the pressure across the
lattice for a desired injection rate and then use that pressure to
solve fluid flow through the network. For two-phase displacement the pressure
across the lattice $\Delta P$ is related to the injection rate $Q$
through the equation
\begin{equation}
\label{eq:Qdarcy}
Q=A\Delta P+B\ .
\end{equation}
Here $A$ and $B$ are parameters depending on the geometry of the
medium and the current configuration of the liquids. The first part of
Eq.\ (\ref{eq:Qdarcy}) is simply Darcy's law for one phase flow, while
the last part $B$ results from the capillary pressure between the two
phases. As long as the menisci do not move $B$ is constant.

The pressure $\Delta p_{ij}$ across each tube can be related to the
pressure across the lattice $\Delta P$. All the equations calculating
the fluid flow in the system, have the functional form $f(x)=ax+b$. As
a consequence $\Delta p_{ij}$ becomes a function of $\Delta P$,
\begin{equation}
\Delta p_{ij}=\Gamma_{ij}\Delta P + \Pi_{ij}\ .
\label{eq:paffine}
\end{equation}
$\Gamma_{ij}$ is a dimensionless quantity depending on the mobilities
($k/\mu_{eff}$) of the tubes and $\Pi_{ij}$ is a
function of the capillary pressures of the menisci inside the
tubes. If no menisci are present $\Pi_{ij}$ is zero.  Eq.\
(\ref{eq:paffine}) can easily be deduced for two cylindrical tubes
with different radii connected in series.

By inserting Eq.\ (\ref{eq:paffine}) into Eq.\ (\ref{eq:tubeflow}) we
obtain after some algebra a relation between the local flow rate
$q_{ij}$ and the pressure $\Delta P$ across the network
\begin{equation}
\label{eq:qdarcy}
q_{ij}=\tilde{a}_{ij}\Delta P+\tilde{b}_{ij}\ .
\end{equation}
The parameter $\tilde{a}_{ij}$ is proportional to $\Gamma_{ij}$ and
the mobility ($k_{ij}/\mu_{eff}$) of the tube $ij$. $\tilde{b}_{ij}$
contains the capillary pressures of the menisci.

The solution due to a constant injection rate can now be summarized
into the following steps: 

1.\ We first find the nodal pressures for two different 
pressures $\Delta P^{'}$ and $\Delta P^{''}$ applied across the 
lattice. 

2.\ From the two solutions of the nodal pressures the
corresponding injection rates $Q^{'}$ and $Q^{''}$ and
the local flow rate $q'_{ij}$ and $q''_{ij}$ are calculated.

3.\ $A$ and $B$ is calculated by solving the two equations obtained
when inserting $\Delta P^{'}$, $Q^{'}$, $\Delta P^{''}$ and $Q^{''}$
into Eq.\ (\ref{eq:Qdarcy}).

4.\ The pressure $\Delta P$ across the lattice for the 
desired $Q$ is then calculated by using Eq.\ (\ref{eq:Qdarcy}).

5.\ This $\Delta P$ is inserted in Eq.\ (\ref{eq:qdarcy}) to get the
local flow $q_{ij}$. Note that the parameters $\tilde{a}_{ij}$ and
$\tilde{b}_{ij}$ must first be found by solving the two equations
obtained by insterting $q'_{ij}$, $\Delta P^{'}$ $q''_{ij}$ and
$\Delta P^{''}$ from step 2 into Eq.\ (\ref{eq:qdarcy}).

\subsection{Moving the menisci}
A time step $\Delta t$ is chosen such that every meniscus is allowed
to travel at most a maximum step length $\Delta x_{max}$ during that
time step. In each time step we check whether or not a meniscus
crosses a node. If this happens, the time step is redefined such that
this meniscus stops at the end of the tube.

A meniscus reaching the end of a tube is moved into the neighbor tubes
according to some defined rules.  These rules take care of the
different fluid arrangements that can appear around the node.
Basically, the non-wetting fluid can either invade into or retreat
from the neighbor tubes as shown in Fig.~\ref{fig:nodeconf}\,(a) and
(b) respectively.  In Fig.~\ref{fig:nodeconf}\,(a) the non-wetting
fluid approaches the node from below (drainage). When the meniscus has
reached the end of the tube (position 1), it is removed and three new
menisci are created at position $\delta$ in the neighbor tubes
(position 2). The distance $\delta$ is about 1--5\% of the tube length
$d$. The small distance $\delta$ avoids that the created menisci at
position 2 immediately disappear and move back to the initial position
1 in tubes where the flow direction is opposite to the direction of
the invading fluid. The total time lapse is adjusted to compensate the
instantaneous change in local volume of the fluids when the menisci
move a distance $\delta $. The time lapse is adjusted such that the
total volume of the fluids always is conserved.

Fig.~\ref{fig:nodeconf}\,(b) shows the opposite case when the
non-wetting fluid retreats into a single tube (imbibition).  As
Fig.~\ref{fig:nodeconf}\,(b) shows the properties of imbibition should
not be neglected as long as the menisci can travel in both
directions. Our approximation in Fig.~\ref{fig:nodeconf}\,(b) cannot
handle important properties found in imbibition such as
film flow and snap off~\cite{Cieplak90,Len83}. However, in drainage
which is what we are focusing on, arrangement (b) will appear rarely
compared to (a). For that reasion, any futher description of imbibition
than the one presented in Fig.~\ref{fig:nodeconf}\,(b), does not
seem necessary.

Summarized, the procedure for each time step $\Delta t$ is:

1.\ The nodal pressures $p_j$ are determined.

2.\ The $p_j$'s are related to the desired injection rate $Q$
from Eqs.\ (\ref{eq:Qdarcy}) and~(\ref{eq:qdarcy}).

3.\ The local flow rate in each tube is computed by using 
Eq.\ (\ref{eq:tubeflow}). 

4.\ The local flow rates are used to calculate the time step $\Delta
t$ such that only one meniscus reaches the end of a tube or travel at
most the step length $\Delta x_{max}$ during that time step.

5.\ The menisci are updated according to $\Delta t$. The total time
lapse is recorded before the whole procedure is repeated for the
new fluid configuration.

\section{Simulations}
\label{sec:result}
In two-phase fluid displacement there are mainly three types of forces:
viscous forces in the invading fluid, viscous forces in the defending
fluid and capillary forces due to the interface between them. This leads
to two dimensionless numbers that characterize the flow: 
the capillary number $C_a$ and the viscosity ratio $M$.

The capillary number describes the competition be\-tween 
capillary and viscous forces. It is defined as
\begin{equation}
C_a=\frac{Q\mu }{\Sigma\gamma }\ ,
\label{eq:Ca}
\end{equation}
where $Q$ ($\mbox{cm}^2/\mbox{s}$) denotes the injection rate, $\mu $
(Poise) is the maximum viscosity of the two fluids , $\Sigma$
($\mbox{cm}^2$) is the cross section of the inlet and $\gamma $
(dyn/cm) is the interfacial tension between the two phases. $\Sigma $
is calculated by taking the product of the length of the inlet and the
mean thickness of the lattice due to the average radius of the tubes.

$M$ defines the ratio of the viscosities of the two fluids and is
given by the invading viscosity $\mu_2$ divided with the defending
viscosity $\mu_1$:
\begin{equation}
M=\frac{\mu_2}{\mu_1}\ .
\end{equation}

In the simulations the pressure across the lattice $\Delta P$ is 
given by Eq.\ (\ref{eq:Qdarcy}) as
\begin{equation}
\Delta P=\frac{Q}{A}-\frac{B}{A}=\frac{Q}{A}+P_{cg}\ .
\label{eq:press}
\end{equation}
Since $B$ is due to the capillary pressure of the menisci, we define
$-B/A$ as the global capillary pressure of the system, $P_{cg}$.
$P_{cg}$ includes the menisci surrounding the trapped cluster of
defending fluid (cluster menisci) as well as the menisci belonging to
the front between the invading and defending fluid (front menisci).

In addition to $P_{cg}$ we calculate \pcfr , the capillary
pressure averaged along the front. \pcfr\ consists only
of the capillary pressures due to the front menisci and we define it as
\begin{equation}
\pcfr = \frac{1}{N}\sum_{\alpha=1}^N \left|p_c^{\alpha}\right|\ .
\label{eq:pcfr}
\end{equation}
Here the index $\alpha$ addresses the tubes in the lattice and in the
summation $\alpha$ runs over all tubes containing a meniscus that
belong to the front. $N$ is the number of such tubes. $p_c^\alpha$ is
the capillary pressure of the front meniscus in tube $\alpha$. See
Fig.~\ref{fig:movie.ip} for an example of a structure obtained after a
simulation with low capillary number. The tubes containing a front
meniscus or belonging to the trapped cluster of defending fluid are
identified by running a Hoshen-Kopelman algorithm~\cite{Stauf92} on
the lattice.

The displacement structure in Fig.~\ref{fig:movie.ip} shows that the
front is a complicated path connecting the left and the right
boundaries of the lattice. The width of the front $w$ is found by
taking the standard deviation of the distances between all the front
tubes and the average position of the front.

For every simulation we have calculated $\Delta P$ and $P_{cg}$ as
functions of time. For some of the simulations we have also computed
the average capillary pressure \pcfr\ along the front and analyzed the
behavior of $A$ in Eq.\ (\ref{eq:Qdarcy}).  As will be discussed below,
$\Delta P$, $P_{cg}$ and \pcfr\ are strongly correlated and $A$ seems
to obey surprisingly simple relations.

We have performed drainage simulations in each of the regimes of
interest: viscous fingering, stable displacement and capillary
fingering with $M=\exf{1.0}{-3}\mbox{, }1.0\mbox{ and } \exf{1.0}{2}$
respectively.  The injection rate was systematically varied for each
of the viscosity ratios. The simulations were performed with
parameters as close as possible to experiments performed
in~\cite{Inge97}.  The length $d$ of all tubes in the lattices were
set equal to $1\,\mbox{mm}$ and the radii $r$ of the tubes were chosen
randomly in the interval $0.05d\leq r\leq d$.  The interfacial tension
was set to $\gamma=30\,\mbox{dyn/cm}$ and the viscosities of the
defending and the invading fluids varied between $0.01\,\mbox{P}$
($\simeq$ water) and $10\,\mbox{P}$ ($\simeq $ glycerol).

\subsection{Viscous fingering, $M<1$}
Our study of this regime consists of a series of simulations with
constant viscosity ratio $M=\exf{1.0}{-3}$. The injection rates and
the capillary number used in the different simulations are listed in
Table~\ref{tabl:visc}. To save computation time, most of the
simulations were performed on a lattice of $25\times 35$
nodes. However, one was performed on a lattice of $60\times 80$ nodes
and the resulting structure of this simulation is shown in
Fig.~\ref{fig:movie.vf}.

In viscous fingering the principal force is due to the viscous
forces in the defending fluid. The pattern formation in 
Fig.~\ref{fig:movie.vf} shows that the invading fluid creates
typical fingers into the defending fluid. The pressure across the
lattice, $\Delta P$, shown in Fig.~\ref{fig:press.vf}, decreases
as the less viscous fluid invades the system. The slope of the average
decreasing pressure function is non-trivial and results from the
fractal development of the fingers.

The global capillary pressure $P_{cg}$, also shown in 
Fig.~\ref{fig:press.vf}, fluctuates around a mean value of about 
$\exf{1}{3}\,\mbox{dyn}/\mbox{cm}^2$. The fluctuation is strongly
correlated in both time and amplitude to the noise in $\Delta P$.
As will be discussed below, the variations correspond to the changes in 
the capillary pressure as a meniscus invades into or retreats from a tube.

We have calculated \pcfr\ for every simulation performed on the lattices
of $25\times 35$ nodes. The result for four of the simulations is shown in
Fig.~\ref{fig:pcapfr-vf} together with the global capillary pressure
$P_{cg}$. The pressures in Fig.~\ref{fig:pcapfr-vf} are normalized
by dividing them with the average threshold pressure of the tubes. The
average threshold pressure is defined as $2\gamma/\langle
r\rangle$ where $\langle r\rangle$ is the mean radius of the tubes.
The mean threshold pressure is about
$\exf{1.1}{3}\,\mbox{dyn}/\mbox{cm}^2$ in all simulations, since the
radii of the tubes always are chosen randomly in the interval
$[0.05,1.0]\,\mbox{mm}$.  Note that in Fig.~\ref{fig:pcapfr-vf},
$P_{cg}$ has been subtracted by $1000\,\mbox{dyn}/\mbox{cm}^2$ before
it was normalized.  This avoids the pressure functions to
overlap at low capillary numbers when they are plotted in the figure.

Fig.~\ref{fig:pcapfr-vf} shows that the fluctuations of \pcfr\
are correlated in time to the fluctuations of $P_{cg}$. 
At lowest $C_a=\exf{3.5}{-4}$, \pcfr\ and $P_{cg}$ are even
indistinguishable. For all simulations except that at lowest 
capillary number the amplitude of the fluctuations in \pcfr\ 
decreases as time increases. For high injection rates \pcfr\ is 
found to approach $1$, which is the mean threshold pressure. 
 
In the simulation at $C_a=\exf{3.5}{-4}$ we approach the regime of
capillary fingering. In slow capillary fingering the local capillary
pressures of the menisci becomes equal when all the menisci are
stable. This is often referred to as capillary equilibrium leading to
$P_{cg}\simeq \pcfr$. However, in fast displacement the local
capillary pressures are generally different. Thus, for a large system
with many menisci the average capillary pressure of the front will
approach the average threshold pressure of the tubes.

Due to the less viscous defending fluid the pressure gradient at high
injection rate is largest at the finger tip closest to the upper
boundary of the lattice. The meniscus in the uppermost finger tip will
therefor more likely continue and invade the next tube compared to the
menisci lying behind it. Thus, the menisci lying behind will get stuck
and the fluctuations in the global capillary pressure can only be
caused by the moving finger tip.  This is seen in
Fig.~\ref{fig:pcaphi} where we have plotted $P_{cg}$ (a) together with
the capillary pressure of the meniscus which is located in the tube
that always has the largest flow rate. The data in
Fig.~\ref{fig:pcaphi} are taken from the simulation performed at
$C_a=\exf{1.1}{-2}$ given in Table~\ref{tabl:visc}.

On basis of Figs.~\ref{fig:pcapfr-vf} and~\ref{fig:pcaphi} we
conclude that the fluctuations in $P_{cg}$ correspond to the local
capillary pressure of the menisci invading the tubes.  Moreover, the
average level of $P_{cg}$ is almost equal \pcfr .

\subsection{Stable displacement, $M>1$}
In the regime of stable displacement we have run seven simulations
spread over different lattices of size between $25\times 35$ and
$60\times 60$ nodes. See Table~\ref{tabl:stable}. The simulations were
run with the viscosity ratio $M=\exf{1.0}{2}$, but at different
capillary numbers. The resulting structure of a simulation performed
on a lattice with $60\times 60$ nodes is visualized in
Fig.~\ref{fig:movie.sd}. The pressure across the lattice $\Delta P$,
the global capillary pressure $P_{cg}$ and $1/A$ were calculated for
all simulations. The results are plotted in Figs.~\ref{fig:press-sd1} 
and~\ref{fig:press-sd2} .

In stable displacements the fluid movements are dominated by the
viscous forces in the invading liquid. The viscous pressure gradient
in the invading phase is found to stabilize the front and a compact
pattern with an almost flat front between the non-wetting and wetting
fluid is generated (see Fig.~\ref{fig:movie.sd}).
The stabilized front introduces a length scale in the
system for large times.  This length scale is identified as the
saturation width $w_s$ of the front~\cite{Inge97}. 

To save computation time the simulations performed on the lattices of
$40\times 40$ and $60\times 60$ nodes (Fig.~\ref{fig:press-sd1}) were
stopped after the width of the front had stabilized. The other
simulations (Fig.~\ref{fig:press-sd2}), however, were run until the
invading fluid reached the outlet.

The pressures in Figs.~\ref{fig:press-sd1} and~\ref{fig:press-sd2}
show a quite different behavior compared to the pressures in
Figs.~\ref{fig:press.vf} and~\ref{fig:pcapfr-vf}.  The pressure across
the lattice $\Delta P$ (a) and the global capillary pressure $P_{cg}$
(b) are both found to increase as the more viscous fluid is injected
into the system.  However, the average slope depend very much on the
injection rate. At low capillary numbers ($C_a=\exf{7.5}{-4}$) we
reach the regime of capillary fingering and similar to viscous
fingering the pressures reduce to that describing the capillary
fluctuations along the front.

To explain the rather unexpected behavior at high injection rates we
have to discuss the effect of the trapped cluster of the defending
fluid left behind the front. In stable displacement the driving
pressure gradient lies between the inlet and the front causing a
pressure drop between the top and the bottom of the trapped
clusters. At moderate injection rates where the clusters stay in place
and keep their shapes, the forces due to the pressure drop must be
canceled by capillary forces acting on the cluster menisci. The sum of
those capillary forces contribute to the observed increase in
$P_{cg}$.

The above interpretation of $P_{cg}$ also applies in the regime of
viscous fingering. In viscous fingering at high capillary numbers
there are few clusters. Thus, the extra contribution to $P_{cg}$ from
the cluster menisci becomes negligible.  When the injection rate is
reduced clusters develop, but now the pressure gradient across the
clusters is small because of the low injection rate.  Therefore, the
contribution to $P_{cg}$ still is negligible. This is in agreement of the
observations in the previous section where $P_{cg}$ was found to
fluctuate around a constant level for both high and low capillary
numbers.

All the displacements listed in Table~\ref{tabl:stable}, except that
at $C_a=\exf{7.5}{-4}$, reached the saturation time $t_s$ where the
front stabilize. In Figs.~\ref{fig:press-sd1} and~\ref{fig:press-sd2}
$t_s$ is indicated by a vertical dashed line. If we neglect the
fluctuations, $\Delta P$ and $P_{cg}$ are approximately linear
functions of time for $t>t_s$.

The linearity in $P_{cg}$ is explained by looking at the generation of
the clusters in the system. We notice that for large times
($t>t_s$), when the front has saturated with fully developed clusters
behind it, the average front position $h$ can only depend on the injection
rate and the viscosity ratio. Both properties are constant through
each simulation, resulting in $h$ being linear in time.  A plot
showing the average front position as a function of time at
$C_a=\exf{4.6}{-3}$ for the lattice with $60\times 60$ nodes is shown
in Fig.~\ref{fig:hmid}.

On average, every cluster contributes to $P_{cg}$ with a certain
amount causing the global capillary pressure to be proportional to the
number of clusters behind the front. For large times, the number of
clusters increases linearly with the average front position causing
$P_{cg}$ to be linear in $h$. Summarized we obtain
\begin{equation}
P_{cg}\propto h\propto t\ \ \ , \ \ \ t>t_s\, .
\label{eq:pcglin}
\end{equation}
It has to be emphasized that there might be large deviations from the
observed linearities when clusters of size comparable to the system
develops. The argument does not apply when $t<t_s$, either. Then the
average change in $\Delta P$ and $P_{cg}$ depends on the fractal
development of the displacement structure.

The quantity $1/A$, defined in Eq.\ (\ref{eq:Qdarcy}), is calculated
by using Eq.\ (\ref{eq:press}).  In Figs.~\ref{fig:press-sd1}
and~\ref{fig:press-sd2} we have plotted $A_0/A$ expressed as
\begin{equation}
\frac{A_0}{A}=A_0\,\frac{\Delta P-P_{cg}}{Q}\ .
\label{eq:aconst}
\end{equation}
Here $A_0$ is equal to the proportionality factor between $Q$ and
$\Delta P$ when only the defending fluid flows trough the lattice,
i.e.\ $A_0=\Sigma K/\mu_1 L$. $K$ denotes the absolute permeability of
the lattice and $L$ is the length of the system. From
Figs.~\ref{fig:press-sd1} and~\ref{fig:press-sd2}, we conclude that at
high capillary numbers and for $t>t_s$, $1/A$ is proportional to the
injection time.

The linearity in $1/A$ is related to the displacement structure that
develop during injection of non-wetting fluid.  Let us return to
equation Eq.\ (\ref{eq:Qdarcy}) which we can rewrite to obtain
\begin{equation}
Q=\Sigma G\left(\Delta P-P_{cg}\right)\frac{1}{L}\ ,
\label{eq:wash1}
\end{equation} 
where $G\equiv AL/\Sigma $ defines the simulated mobility of the
system. It must be emphasized that this mobility may be hard to measure
experimentally and it should not be confused with the conventional
effective mobility of the system, denoted by $G_{eff}$ (see below).
Eq.\ (\ref{eq:wash1}) looks very similar to Eq.\ (\ref{eq:tubeflow})
describing the fluid flow through a single tube. If we now assume that
an equation of the form~(\ref{eq:tubeflow}) applies on the whole lattice
we interpret $G$ as
\begin{equation}
G=\frac{K}{\mu_{eff}}\ .
\label{eq:wash2}
\end{equation} 
Here $\mu_{eff}$ is the effective viscosity of the whole system.
$\mu_{eff}$ is not trivial to calculate because of the fractal
structure of the trapped clusters and front.  However, if we denote
the normalized average front position as $\hat{h}$ and assume that the
viscosity of the region behind the front can be expressed by an
effective viscosity ${\mu_2}'$, $\mu_{eff}$ becomes
\begin{equation}
\mu_{eff}=(1-\hat{h})\mu_1+\hat{h}{\mu_2}'\ .
\label{eq:vieff}
\end{equation}
Here $0\!<\!\hat{h}\!<\!1$. By combining Eqs.\ (\ref{eq:vieff}) and
(\ref{eq:wash2}), we finally obtain after some reorganization
\begin{equation}
\frac{1}{G}\propto \frac{1}{A}=\frac{1}{A_0}\left[(M_e-1)\hat{h}+1\right]\ ,
\label{eq:ainv}
\end{equation}
where $M_e\equiv{\mu_2}'/\mu_1$ defines the effective viscosity ratio.

To a first approximation, we have calculated ${\mu_2}'$ directly from
the displacement structure by taking the sum of the fluid viscosities
multiplied by their respective macroscopic saturations behind the
front.  However, the calculated ${\mu_2}'$ was not consistent with
$(M_e-1)$ given by the average slope of $A_0/A$ plotted in
Figs.~\ref{fig:press-sd1} and~\ref{fig:press-sd2}. This indicates that
${\mu_2}'$ not only depend on the fluid saturations but also the
structure of the liquids behind the front. Thus, microscopic
properties of the fluid configurations has to be considered in order
to find the correct $M_e$.

We are now in a position to derive a more exact formalism describing
the pressure evolution and the changes in the effective mobility
$G_{eff}$ of the system when clusters develop. $G_{eff}$ corresponds
to the ratio of the effective permeability and the effective
viscosity of the system. If we again look at the global capillary
pressure $P_{cg}$, it can be interpreted as a sum of the capillary
pressure along the front menisci, $P_{mf}$ and the capillary pressure
of the cluster menisci, $P_{mc}$. Furthermore, for large times
($t>t_s$) we can express $P_{mc}=\Delta_{mc}\, h$ giving
\begin{equation}
P_{cg}=\Delta_{mc}\,h+P_{mf}\ .
\label{eq:globalpc}
\end{equation}
Here $\Delta_{mc}$ denotes the proportionality factor between $P_{cg}$
and $h$ where $0\!<\!h\!<\!L$ is the average front position.  In the
limit of very low injection rate the capillary pressure of the front
and cluster menisci will approach capillary equilibrium causing
$\Delta_{mc} \rightarrow 0$ and $P_{mf}\simeq\pcfr\simeq P_{cg} $.

We can use the relation for $P_{cg}$ in Eq.\ (\ref{eq:globalpc}) to
deduce a formula for $G_{eff}$. What we are seeking is an equation for
the flow rate $U$, on the form of Darcy's law
\begin{equation}
U=G_{eff}\left(\Delta P-P_{mf}\right)\frac{1}{L}\ .
\label{eq:G1}
\end{equation} 
The pressure gradient $(\Delta P-P_{mf})/L$ accounts for the capillary
pressure due to the front, $P_{mf}$. 

We start by inserting Eq.\ (\ref{eq:globalpc}) into 
Eq.\ (\ref{eq:wash1}) to get
\begin{equation}
U=\frac{K}{\mu_{eff}}\left(\Delta P-\Delta_{mc}\, h-P_{mf}\right)\frac{1}{L}\ ,
\label{eq:U1}
\end{equation}
where the flow velocity $U\equiv Q/\Sigma$.  Eq.\ (\ref{eq:U1}) can be
written into the from of Darcy's to obtain
\begin{equation}
U=\frac{K}{\mu_{eff}+\mu_{mc}}\left(\Delta P-P_{mf}\right)\frac{1}{L}\ ,
\label{eq:U2}
\end{equation}
where
\begin{equation}
\mu_{mc}\equiv\frac{K}{UL}\,\Delta_{mc}\, h\ .
\label{eq:mumc}
\end{equation}
From Eq.\ (\ref{eq:U2}) we immediately interpret $\mu_{mc}$ as the
increase in viscosity of the invaded region caused by the cluster
formation. However, this may just as well be seen as a decrease in
the effective permeability behind the front~\cite{Inge97}.  Note that the $U$
dependency in Eq.\ (\ref{eq:mumc}) only indicates changes in
$\Delta_{mc}$ between displacements executed at different
injection rates. The behavior when the flow rate changes during one
displacement is not discussed here. From Eq.\ (\ref{eq:U2}) we
finally define the effective mobility of the lattice as
\begin{equation}
G_{eff}\equiv \frac{K}{\mu_{eff}+\mu_{mc}}\ .
\label{eq:G2}
\end{equation}

There is one important interpretation of $G_{eff}$ when the average
front position has reached the outlet, i.e. $h=L$. Then only the
invading fluid flows through the system and $G_{eff}$ becomes equal to
the effective mobility of the invading phase. If we insert Eqs.\
(\ref{eq:vieff}) and (\ref{eq:mumc}) into Eq.\ (\ref{eq:G2}) we obtain
when $h=L$
\begin{equation}
G_{eff}(h=L)=\frac{K}{{\mu_2}'+\frac{K\Delta_{mc}}{U}}\ .
\label{eq:G3}
\end{equation}
From this expression follows directly a relation for the relative
permeability of the invading phase, $k_{ri}$ defined as
\begin{equation}
k_{ri}=\frac{G_{eff}\, (h=L){\mu_2}'}{K}\ .
\label{eq:G4}
\end{equation}
Note that ${\mu_2}'=\mu_{eff}(\hat{h}=1)$ from Eq.\ (\ref{eq:vieff}) and
that ${\mu_2}'$ is given by the slope of $1/A$ in Eq.\ (\ref{eq:ainv}). 

The above formalism resulting in Eq.\ (\ref{eq:G2}), takes into
account the capillary pressure of the cluster menisci as well as the
capillary pressure along the front.  However, at moderate injection
rates where the clusters stay in place and keep their shapes, we have
found $G_{eff}$ directly in the simulations by assigning zero
permeability to tubes belonging to the trapped clusters.  Thus, the
clusters will be frozen to their initial positions and in the
calculations they are treated as additional boundary conditions where
fluid cannot flow.  Simulations have provided evidence that when the
clusters are frozen the parameter $A$ in Eq.\ (\ref{eq:Qdarcy})
adjusts such that the simulated $G$ in Eq.\ (\ref{eq:wash1}) becomes
equal to the calculated $G_{eff}$ in Eq.\ (\ref{eq:G2}). $G_{eff}$ is
calculated by using $P_{cg}$ and $\Delta P$ which are plotted in
Figs.~\ref{fig:press-sd1} and~\ref{fig:press-sd2}. Moreover, with
frozen clusters the global capillary pressure $P_{cg}$, reduces to
that describing the capillary pressure along the front, $P_{mf}$.

\subsection{Capillary fingering, $M=1$}
In the regime of capillary fingering we have run 17 simulations with
viscosity matching fluids ($M=1.0$) spread over six different capillary
numbers. The different capillary numbers and the corresponding
injection rates are listed in Table~\ref{tabl:invper}. The lattice
size was $40\times 60$ nodes for all simulations. Due to long
computation time we only did two simulations at the lowest
capillary number. For all the other capillary numbers, we ran three
simulations.

In capillary fingering the displacement is so slow that the viscous
forces are negligible, with the consequence that the main force is the
capillary one between the two fluids.  Only the strength of the
threshold pressure in a given tube decides whether the invading fluid 
invades that tube or not. Since the radii of the tubes (which determine
the threshold pressures) are randomly chosen from a given
interval, the non-wetting fluid flows along the path of least
resistance. 

The displacement structure of one of the simulations at lowest
$C_a=\exf{4.6}{-5}$ is shown in Fig.~\ref{fig:movie.ip}. We observe
that the invading fluid creates a  rough front with
trapped clusters that appear at all scales between the tube 
length and the maximum width of the front.

For all simulations $\Delta P$ and $P_{cg}$ were calculated. The
result for six of the simulations each at one of the different
capillary numbers are shown in Fig.~\ref{fig:press-ip}. In the
figure (a) denotes $\Delta P$ and (b) denotes $P_{cg}$. Note that
$P_{cg}$ has been subtracted by $1000\,\mbox{dyn}/\mbox{cm}^2$ to
avoid overlap of the curves at low capillary numbers.

The front was found to stabilize at high injection
rates~\cite{Inge97} and the saturation time $t_s$ is indicated by the
vertical dashed line in Fig.~\ref{fig:press-ip}. At the lowest
capillary number it might be difficult to estimate the 
saturation time accurately. At very low injection rate the 
width of the front probably approaches an upper cut off equal to
the finite size of the lattice.

At high capillary numbers for $t>t_s$, $\Delta P$ and
$P_{cg}$ are found to increase linearly as a function of time. This is
consistent with the result from stable displacement. However, the
difference $\Delta P-P_{cg}$, is constant through each of the
simulation opposed to what we observed when $M>1$.
Fig.~\ref{fig:aconst} shows the difference $\Delta P-P_{cg}$ for one
of the simulations at $C_a=\exf{2.3}{-4}$. The plot shows the
normalized value of $\Delta P-P_{cg}$ subtracted by $1$ such that the
resulting data fall close to zero. The fluctuations appearing in the
$9$th digit are caused by numerical round off errors in the simulations.

In the previous section we suggested that the mobility $G$ defined in
Eq.\ (\ref{eq:wash2}), was a function of the effective viscosity
$\mu_{eff}$. Moreover, we assumed that $\mu_{eff}$ depends on the
saturation of the invading and the defending fluid and the
displacement structure.  When $\mu_1=\mu_2$ none of these assumptions
apply. From Eq.\ (\ref{eq:wash1}) the constant pressure difference
$\Delta P-P_{cg}$ in Fig.~\ref{fig:aconst} implies that the simulated
mobility $G$ becomes constant, even with respect to the local
displacement structure.  The effective viscosity reduces to the
viscosity of the two liquids, $\mu_{eff}=\mu_1=\mu_2$ giving
$A=A_0=\Sigma K/\mu_1 L$.  The difference $\Delta P-P_{cg}$, however,
depend on the injection rate $Q$. This is observed in
Fig.~\ref{fig:press-ip} where $\Delta P-P_{cg}$ increases for
increasing capillary number.

When $\mu_1=\mu_2$ the pressure difference $\Delta P-P_{cg}$ has a
simple interpretation. It corresponds to the viscous pressure drop
arising when viscous fluids are moving. When the injection rate is
reduced the viscous pressure drop becomes negligible and we approach
the regime of capillary fingering. At low injection rates the pressure
gradient across the trapped clusters also vanish giving only a small
increase in $\Delta P$ and $P_{cg}$.  This is observed in
Fig.~\ref{fig:press-ip} at $C_a=\exf{4.6}{-5}$ where the average
increase in $P_{cg}$ becomes quite small and $\Delta P\simeq
P_{cg}$. In the same figure, the sudden jumps in the pressure function
identify the bursts where the invading fluid proceeds abruptly. This
corresponds to Haines jumps~\cite{Hain30,Maloy92}.

The property that the mobility $G$ is constant when the liquids
have equal viscosities simplifies the computation of the nodal
pressures in the lattice. By substituting $A$ with $A_0$ in
Eq.\ (\ref{eq:Qdarcy}) we find that the injection rate is given by
\begin{equation}
Q=A_0\Delta P + B\ .
\label{eq:Qdarcyeq}
\end{equation}
This equation has only one unknown, the term $B$, opposed to the
original Eq.\ (\ref{eq:Qdarcy}) having two unknowns, both $A$ and $B$.
To verify the result when $A$ is replaced by $A_0$ we have compared
the solution found from Eq.\ (\ref{eq:Qdarcyeq}), necessitating one
solution of the flow equations, with the one given when Eq.\
(\ref{eq:Qdarcy}) is solved twice. Not surprisingly, there is
excellent agreement between these two results.

The strong evidence that $A=A_0$ is constant in Eq.\ (\ref{eq:Qdarcyeq})
can be deduced from simple considerations of the energy dissipation
in the system. In analogy with electrical circuits, we define
the total energy dissipation $W$ in the system as
\begin{equation}
W=Q\Delta P\ .
\label{eq:diss1}
\end{equation}

The total dissipation must equals the sum of the dissipation in
every tube $\alpha$ in the lattice. Thus,
\begin{equation}
Q\Delta P=\sum_{\alpha} q_{\alpha}\Delta p_{\alpha}\ ,
\label{eq:diss2}
\end{equation}
where the summation index $\alpha$ runs over all tubes in the lattice.
$q_{\alpha}$ is given by Eq.\ (\ref{eq:tubeflow}) which we rewrite as 
($ij\rightarrow\alpha$)
\begin{equation}
q_{\alpha}=a_{\alpha}\Delta p_{\alpha} + b_{\alpha}\ .
\label{eq:diss3}
\end{equation}
Here we note that $a_{\alpha}$ is proportional to $k_{\alpha}/\mu_{eff}$, the
mobility of the tube $\alpha$. By inserting Eq.\ (\ref{eq:Qdarcy})
and~(\ref{eq:diss3}) into Eq.\ (\ref{eq:diss2}) we get after some
reorganization
\begin{equation}
A\Delta P +B=\sum_{\alpha} a_{\alpha}\Delta P\left(\frac{\Delta p_{\alpha}}{\Delta P}\right)^2+b_{\alpha}\frac{\Delta p_{\alpha}}{\Delta P}\ .
\label{eq:diss4}
\end{equation}
By replacing the local pressure $\Delta p_{\alpha}$ in Eq.\
(\ref{eq:diss4}) with $\Delta P$, using Eq.\ (\ref{eq:paffine}) we
obtain after some algebra
\begin{eqnarray}
Q=&&\left[\sum_{\alpha} a_{\alpha}{\Gamma_{\alpha}}^2\right]\Delta P
   +\left[\sum_{\alpha} \Gamma_{\alpha}(2a_{\alpha}\Pi_{\alpha}+
	b_{\alpha})\right.\nonumber \\
  &&+\left.\frac{\Pi_{\alpha}}{\Delta P}(a_{\alpha}\Pi_{\alpha}+
	b_{\alpha})\right]\ .
\label{eq:diss6}
\end{eqnarray}
If we compare the above equation with Eq.\ (\ref{eq:Qdarcy}) we
recognize the first summation as $A$ and the second as $B$. Thus, $A$
depends entirely on $a_{\alpha}$ and $\Gamma_{\alpha}$.  As stated
earlier, both $a_{\alpha}$ and $\Gamma_{\alpha}$ are proportional to
the mobility of the tubes.  The mobility of each tube depends on the
local fluid configurations, through the effective viscosity
$\mu_{eff}$ as defined in Eq.\ (\ref{eq:vieff}). However, when the
fluids have equal viscosities we get $\mu_{eff}=\mu_1=\mu_2$. As a
consequence $a_{\alpha}$ and $\Gamma_{\alpha}$ becomes constant, which
is consistent with the simulation result.

\section{Discussions}
\label{sec:discuss}
We have simulated drainage displacements at different injection rates
for three different viscosity ratios $M=\exf{1.0}{-3},\ 1.0\mbox{ and
}\exf{1.0}{2}$. The main focus of the work is the study of the
temporal evolution of the pressure when a non-wetting fluid displaces
a wetting fluid in porous media. Moreover, the effect of the trapped
clusters on the displacement process has been discussed. From the
results we clearly see that the capillary forces play an important role
at both high and low injection rates.

At high injection rates with $M\ge 1.0$ the global capillary pressure,
$P_{cg}$ was found to increase as a function of the number of trapped
clusters behind the front. For large times when the front has
saturated, $P_{cg}$ becomes even proportion to the average front
position $h$. This lead to a formalism describing the evolution of the
effective mobility when the non-wetting fluid was injected into the
system. Moreover, we showed that this effective mobility could be used to
estimate the relative permeability of the invading phase when the
average front position has reached the outlet.

At moderate injection rates the effective mobility given by
$P_{cg}$, were shown to be equivalent to assigning zero permeability
to tubes belonging to the trapped clusters.  When $M\ll 1.0$ or at low
injection rates the effect of the clusters become negligible reducing
$P_{cg}$ to describe the local capillary fluctuations of the invading
menisci along the front.  With displacements performed with equal
viscosities, $M=1.0$, we found that the difference $\Delta P-P_{cg}$
was constant. This was shown to be consistent with the energy
dissipation in the system.

It must be emphasized that the properties we report are only valid
for drainage. So far the model is not capable to simulate imbibition.
The simulations have also been performed on a two-dimensional porous
system where clusters develop more easily,  compared to fluid flow in
three-dimensional porous media~\cite{Wilk86}. Moreover, the lattice sizes
are limited by the computation time and more sophisticated and
efficient algorithms have to be developed in order to increase the
system sizes and thereby improve the above results. Another, and
not less important exercise is to compare our simulation results
with experimental measurements.

\acknowledgments The authors thank S.\ Basak, G.\ G.\ Batrouni, E.\
G.\ Flekk\o y and J.\ Schmittbuhl for valuable comments.  The work is
supported by NFR, the Norwegian Research Council, through a ``SUP''
program and a grant of computation time. The work has also got
computer time from the Idris in Paris and from HLRZ,
For\-schungs\-zent\-rum J\"{u}\-lich GmbH.



\begin{figure}
\caption{Flow in a tube containing a meniscus.}
\label{fig:tube}
\end{figure}

\begin{figure}
\caption{Four different fluid arrangements inside one tube. The shaded
and the white regions indicate the non-wetting and wetting fluid 
respectively.}
\label{fig:tubeconf}
\end{figure}

\begin{figure}
\caption{The motion of the menisci at the nodes. (a) The non-wetting
fluid (shaded) reaches the end of the tube (position 1) and is moved a
distance $\delta$ into the neighbor tubes (position 2). (b) The
wetting fluid (white) reaches the end of the tubes (position 1) and
the non-wetting fluid (shaded) retreat to position 2.  To conserve the
volume of the fluids a proper time is recorded due to the small 
movement $\delta $ in (a) and (b).}
\label{fig:nodeconf}
\end{figure}

\begin{figure}
\caption{The displacement structure obtained from a simulation
at $C_a=\exf{4.6}{-5}$ and $M=1.0$. The size of the lattice is
$40\times 60$ nodes and the invading non-wetting fluid (black)
displaces the defending wetting fluid (gray) from below. Notice the
rough front between the fluids and the trapped cluster of defending fluid 
left behind.}
\label{fig:movie.ip}
\end{figure}

\begin{figure}
\caption{The displacement structure obtained by a simulation in the
regime of viscous fingering on a lattice of $60\times 80$ nodes.
$C_a=\exf{4.6}{-3}$ and $M=\exf{1.0}{-3}$. The invading, non-wetting
fluid (black) displaces the defending, wetting fluid (gray) from
below.}
\label{fig:movie.vf}
\end{figure}

\begin{figure}
\caption{$\Delta P$ (a) and $P_{cg}$ (b) plotted as a function of time.
$C_a=\exf{4.6}{-3}$ and $M=\exf{1.0}{-3}$.}
\label{fig:press.vf}
\end{figure}

\begin{figure}
\caption{\pcfr\ (a), and $P_{cg}$ (b), at four different capillary
numbers for the simulations with $M=\exf{1.0}{-3}$ at lattice size
$25\times 35$ nodes. The pressures are normalized using the
average threshold pressure of the tubes. Note that $P_{cg}$ has been
subtracted by $1000\,\mbox{dyn}/\mbox{cm}^2$ before it was normalized
to avoid overlap between the two curves at low capillary numbers.}
\label{fig:pcapfr-vf}
\end{figure}

\begin{figure}
\caption{$P_{cg}$ (a) and the capillary pressure of the meniscus
traveling with the highest velocity (b) at $C_a=\exf{1.1}{-2}$ and
$M=\exf{1.0}{-3}$ in the time interval between $7.0$ and
$12.0\,\mbox{s}$.}
\label{fig:pcaphi}
\end{figure}

\begin{figure}
\caption{The displacement structure obtained of a simulation in the
regime of stable displacement on a lattice of $60\times 60$ nodes.
$C_a=\exf{4.6}{-3}$ and $M=\exf{1.0}{2}$. The invading, non-wetting
fluid (black) displaces the defending, wetting fluid (gray) from
below.}
\label{fig:movie.sd}
\end{figure}

\begin{figure}
\caption{$\Delta P$ (a), $P_{cg}$ (b) and $A_0/A$ (c) as functions of
time for the simulations with $M=\exf{1.0}{2}$ at lattice size
$60\times 60$ (to the left) and $40\times 40$ nodes (to the
right). The vertical dashed lines indicate the saturation time where
the front stabilize.}
\label{fig:press-sd1}
\end{figure}

\begin{figure}
\caption{$\Delta P$ (a), $P_{cg}$ (b) and $A_0/A$ (c) as functions of
time for the five simulations with $M=\exf{1.0}{2}$ at lattice size
$25\times 35$ nodes. Note that for $C_a=\exf{4.2}{-3}\mbox{,
}\exf{2.2}{-3}\mbox{ and } \exf{7.5}{-4}$ $P_{cg}$ has been subtracted
by $1000\,\mbox{dyn}/\mbox{cm}^2$ to avoid overlap of the curves. The
vertical dashed lines indicate the saturation time where the front
stabilize.}
\label{fig:press-sd2}
\end{figure}

\begin{figure}
\caption{The average front position as a function of time at
$C_a=\exf{4.6}{-3}$ and $M=\exf{1.0}{2}$ for the lattice of $60\times
60$ nodes.}
\label{fig:hmid}
\end{figure}

\begin{figure}
\caption{$\Delta P$ (a) and $P_{cg}$ (b) for six simulations with
$M=1.0$ each at one of the capillary number listed in
Table~\ref{tabl:invper}.  The dashed line indicates the saturation
time $t_s$ when the front has reached the saturation width $w_s$. Note
that $P_{cg}$ has been subtracted by $1000\,\mbox{dyn}/\mbox{cm}^2$ to
avoid the curves from overlapping at low capillary numbers.}
\label{fig:press-ip}
\end{figure}

\begin{figure}
\caption{The normalized difference $\Delta P-P_{cg}$ subtracted by 1 at
$C_a=\exf{2.3}{-3}$ and $M=1.0$.}
\label{fig:aconst}
\end{figure}


\begin{table}
\caption{The lattice size and the values for the injection rate and
the capillary number when $M=\exf{1.0}{-3}$.}
\label{tabl:visc}
\begin{tabular}{cdc}
Size & Injection rate  & $C_a$ \\ 
(nodes) & ($\mbox{cm}^3/\mbox{min}$) &  \\ \hline 
$60\times 80$ & 1.5   &  $\exf{4.6}{-3}$ \\
$25\times 35$ & 1.4   &  $\exf{1.1}{-2}$ \\
$25\times 35$ & 0.98  &  $\exf{7.1}{-3}$ \\
$25\times 35$ & 0.62  &  $\exf{4.7}{-3}$ \\
$25\times 35$ & 0.50  &  $\exf{3.6}{-3}$ \\
$25\times 35$ & 0.099 &  $\exf{7.2}{-4}$ \\
$25\times 35$ & 0.049 &  $\exf{3.5}{-4}$ \\
\end{tabular}
\end{table}

\begin{table}
\caption{The lattice size and the values for the injection rate and
the capillary number when $M=\exf{1.0}{2}$.}
\label{tabl:stable}
\begin{tabular}{cdc}
Size & Injection rate  & $C_a$ \\ 
(nodes) & ($\mbox{cm}^3/\mbox{min}$) &  \\ \hline 
$60\times 60$ & 1.5    &  $\exf{4.6}{-3}$ \\
$40\times 40$ & 1.0    &  $\exf{4.6}{-3}$ \\
$25\times 35$ & 2.5    &  $\exf{1.8}{-2}$ \\
$25\times 35$ & 1.3    &  $\exf{9.5}{-3}$ \\
$25\times 35$ & 0.57   &  $\exf{4.2}{-3}$ \\
$25\times 35$ & 0.29   &  $\exf{2.2}{-3}$ \\
$25\times 35$ & 0.10   &  $\exf{7.5}{-4}$ \\
\end{tabular}
\end{table}

\begin{table}
\caption{The values for the injection rate and
the capillary number when $M=1.0$. The lattice is $40\times 60$ nodes.}
\label{tabl:invper}
\begin{tabular}{cdc}
Runs   & Injection rate  & $C_a$ \\ 
      & ($\mbox{cm}^3/\mbox{min}$) &  \\ \hline 
 3    & 10     & $\exf{2.3}{-3}$   \\
 3    & 4.0    & $\exf{9.2}{-4}$   \\
 3    & 2.0    & $\exf{4.6}{-4}$   \\
 3    & 1.0    & $\exf{2.3}{-4}$   \\
 3    & 0.40   & $\exf{9.2}{-5}$   \\
 2    & 0.20   & $\exf{4.6}{-5}$  
\end{tabular}
\end{table}


\begin{thebibliography}{10}

\bibitem{Maloy85}
K.~J. M{\aa}l{\o}y, J.~Feder, and T.~J{\o}ssang, {\em Phys. Rev. Lett.} {\bf
  55}, 26881 (1985).

\bibitem{Chen-Wilk85}
J.-D. Chen and D.~Wilkinson, {\em Phys. Rev. Lett.} {\bf 55}, 1892 (1985).

\bibitem{Len88}
R.~Lenormand, E.~Touboul, and C.~Zarcone, {\em J. Fluid Mech.} {\bf 189}, 165
  (1988).

\bibitem{Cieplak88}
M.~Cieplak and M.~O. Robbins, {\em Phys. Rev. Lett.} {\bf 60}, 2042 (1988).

\bibitem{Cieplak90}
M.~Cieplak and M.~O. Robbins, {\em Phys. Rev. B.} {\bf 41}, 11508 (1990).

\bibitem{Cieplak91}
N.~Martys, M.~Cieplak, and M.~O. Robbins, {\em Phys. Rev. Lett.} {\bf 66}, 1058
  (1991).

\bibitem{Len85}
R.~Lenormand and C.~Zarcone, {\em Phys. Rev. Lett.} {\bf 54}, 2226 (1985).

\bibitem{Witten81}
T.~A. Witten and L.~M. Sander, {\em Phys. Rev. Lett.} {\bf 47}, 1400 (1981).

\bibitem{Pater84}
L.~Paterson, {\em Phys. Rev. Lett.} {\bf 52}, 1621 (1984).

\bibitem{Wilk83}
D.~Wilkinson and J.~F. Willemsen, {\em J. Phys. A} {\bf 16}, 3365 (1983).

\bibitem{Kop-Lass85}
J.~Koplik and T.~J. Lasseter, {\em SPEJ} {\bf 22}, 89 (1985).

\bibitem{Paya86-1}
M.~M. Dias and A.~C. Payatakes, {\em J. Fluid Mech.} {\bf 164}, 305 (1986).

\bibitem{King87}
P.~R. King, {\em J. Phys. A} {\bf 20}, L529 (1987).

\bibitem{Blunt90}
M.~Blunt and P.~King, {\em Phys. Rev. A} {\bf 42}, 4780 (1990).

\bibitem{Blunt91}
M.~Blunt and P.~King, {\em Transp. Porous Media} {\bf 6}, 407 (1991).

\bibitem{Reeves96}
P.~C. Reeves and M.~A. Celia, {\em Water Resour. Res.} {\bf 32}, 2345 (1996).

\bibitem{Paya96}
G.~N. Constantinides and A.~C. Payatakes, {\em AIChE Journal} {\bf 42}, 369
  (1996).

\bibitem{Oeren96}
E.~W. Pereira, W.~V. Pinczewski, D.~Y.~C. Chan, L.~Paterson, and P.~E. {\O}ren,
  {\em Transp. Porous Media} {\bf 24}, 167 (1996).

\bibitem{Blunt97}
D.~H. Fenwick and M.~J. Blunt.
  {\em SPE 38881, proc. of the SPE Annual Tech. Conf.}, San Antonio,
  Texas, U.S.A., Oct. 1997.

\bibitem{Aker97}
E.~Aker, K.~J. M{\aa}l{\o}y, A.~Hansen, and G.~G. Batrouni,
\newblock ``A two-dimensional network simulator for two-phase flow in porous
  media''.
\newblock Submitted to {\em Transp. Porous Media}, 1997.

\bibitem{Wash21}
E.~W. Washburn, {\em Phys. Rev.} {\bf 17}, 273 (1921).

\bibitem{CGM}
G.~G. Batrouni and A.~Hansen, {\em J. Stat. Phys.} {\bf 52}, 747 (1988).

\bibitem{Len83}
R.~Lenormand, C.~Zarcone, and A.~Sarr, {\em J. Fluid. Mech.} {\bf 135}, 337
  (1983).

\bibitem{Stauf92}
D.~Stauffer and A.~Aharony.
\newblock {\em Introduction to percolation theory}.
\newblock Taylor \& Francis, London, Great Britain, 1992.

\bibitem{Inge97}
O.~I. Frette, K.~J. M{\aa}l{\o}y, J.~Schmittbuhl, and A.~Hansen, {\em Phys.
  Rev. E.} {\bf 55}, 2969 (1997).

\bibitem{Hain30}
W.~B. Haines, {\em J. Agr. Sci.} {\bf 20}, 97 (1930).

\bibitem{Maloy92}
K.~J. M{\aa}l{\o}y, L.~Furuberg, J.~Feder, and T.~J{\o}ssang, {\em Phys. Rev.
  Lett.} {\bf 68}, 2161 (1992).

\bibitem{Wilk86}
D.~Wilkinson, {\em Phys. Rev. A} {\bf 34}, 1380 (1986).

\end{thebibliography}
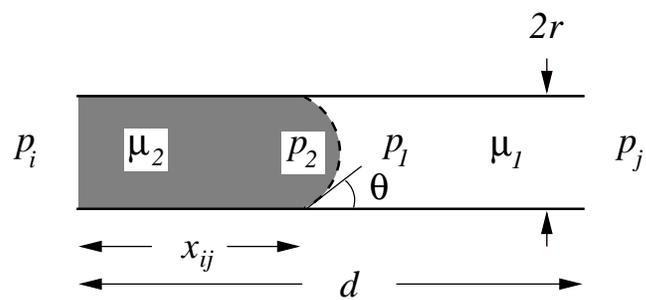
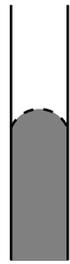
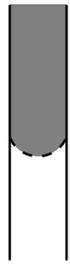
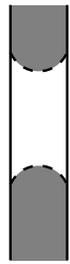
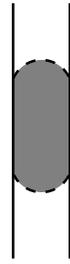
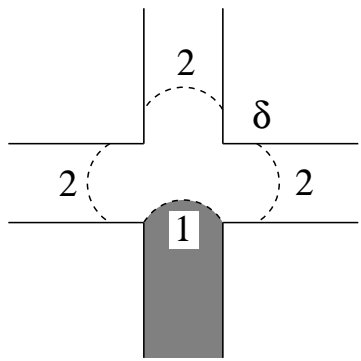
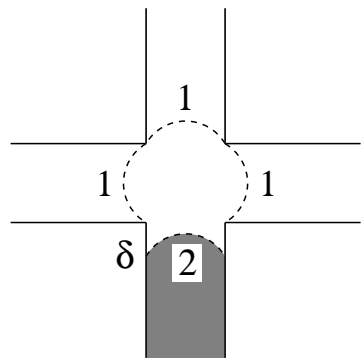
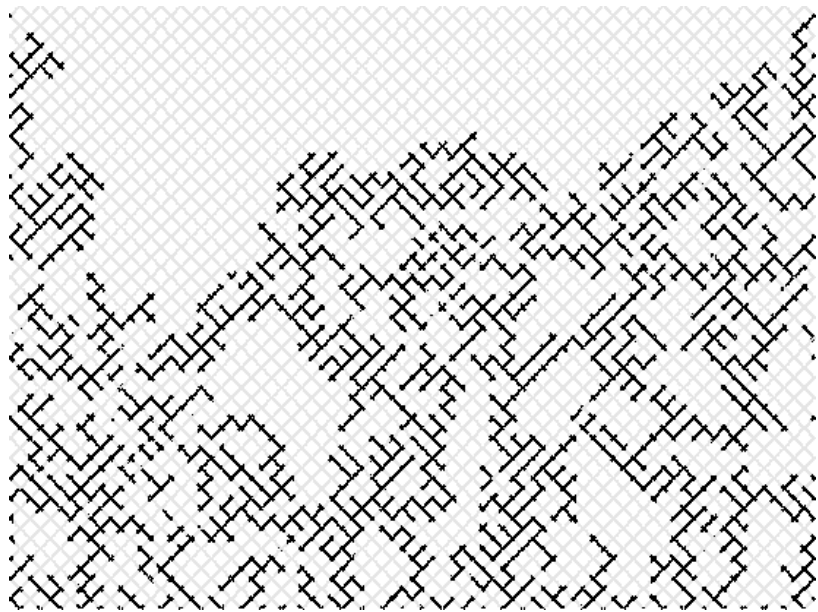
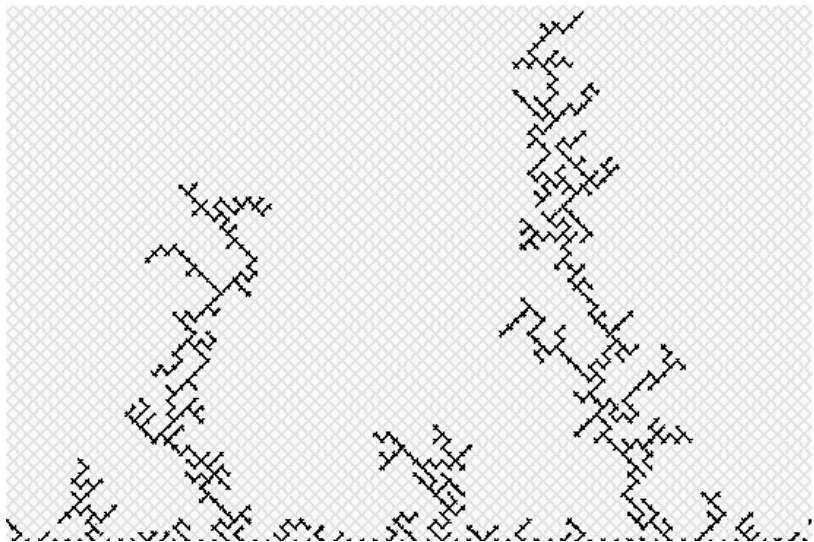
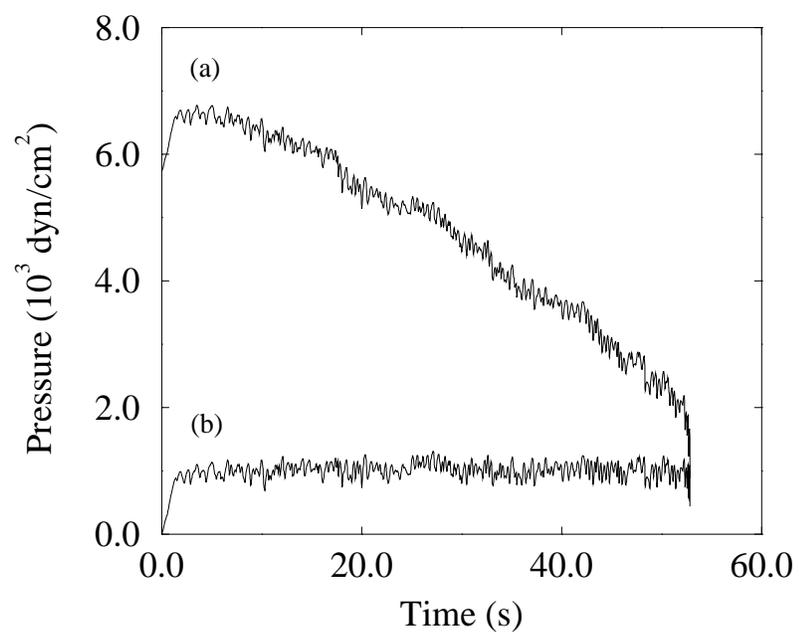
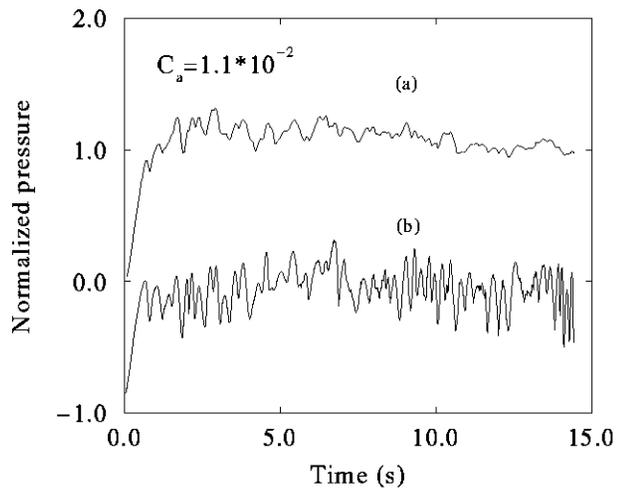
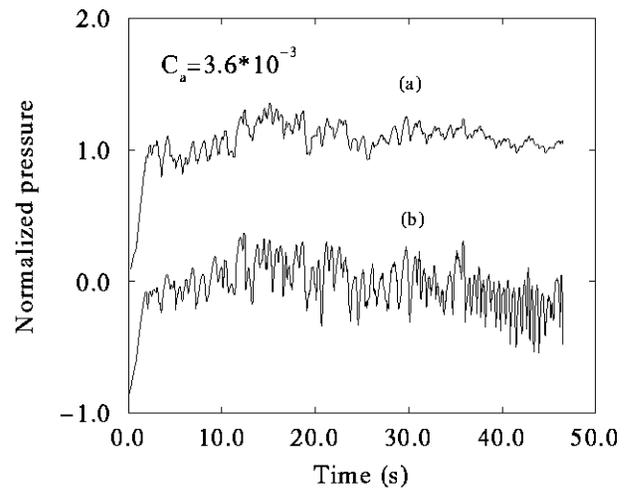
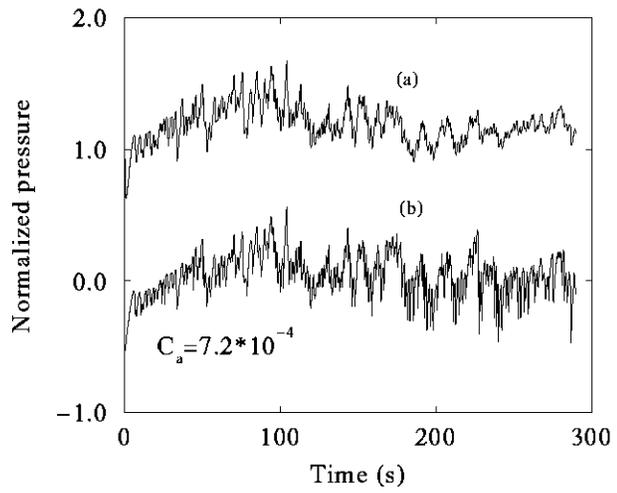
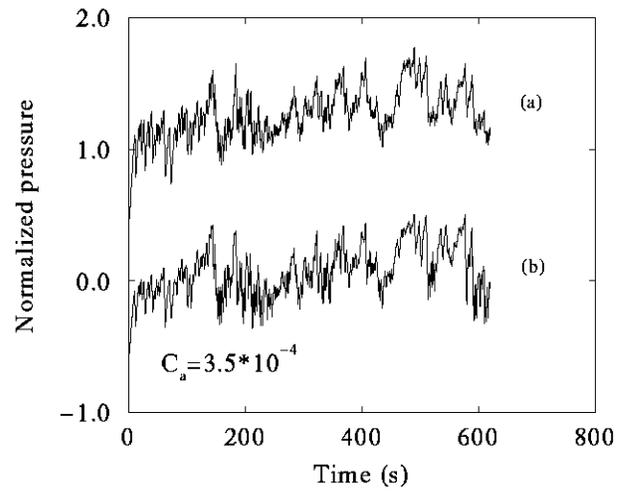
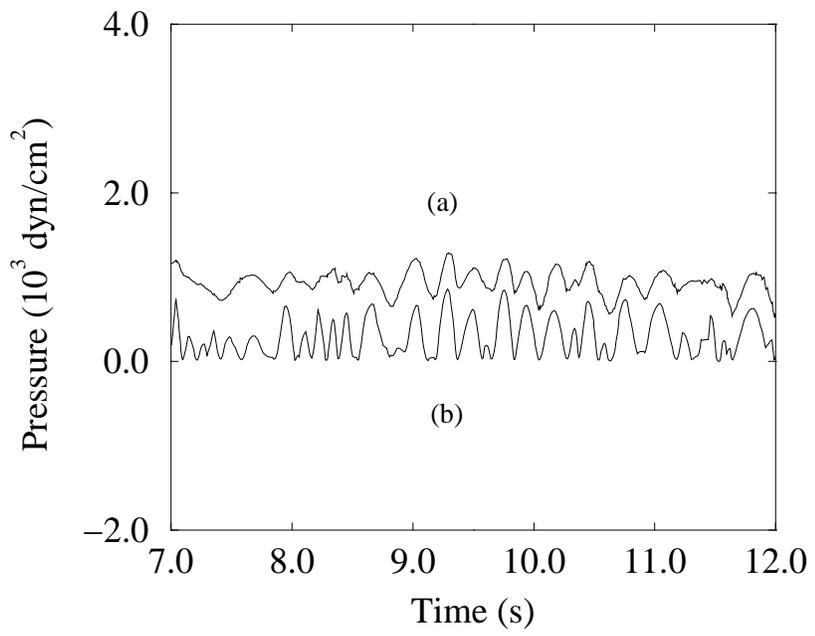
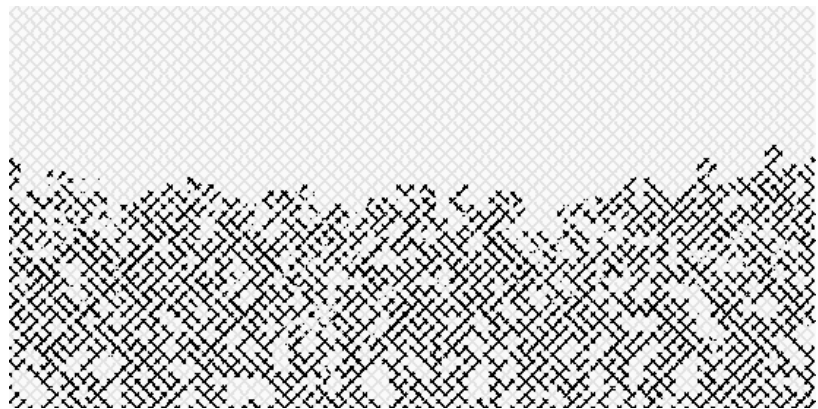
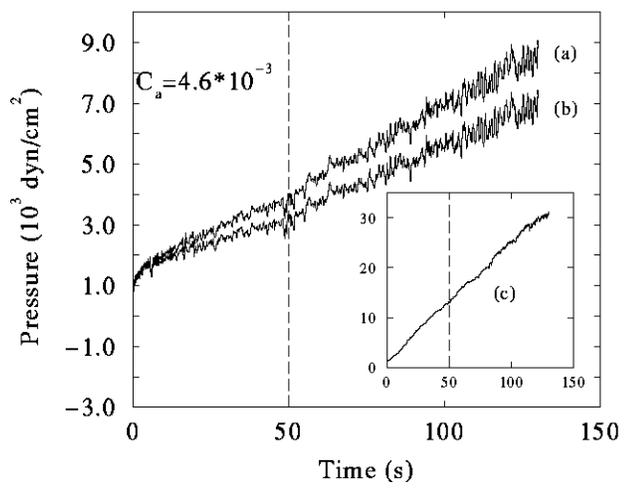
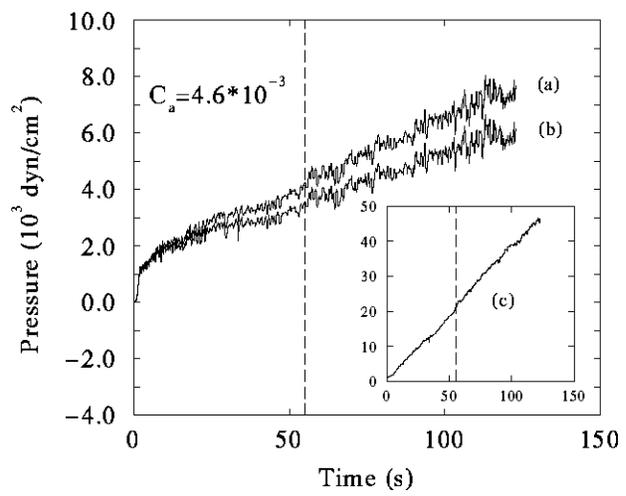
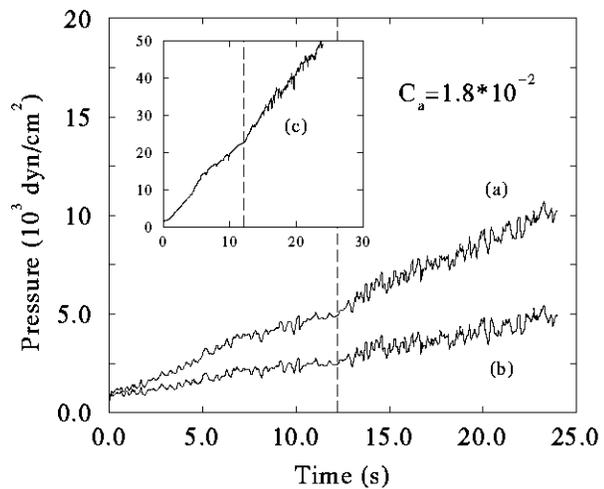
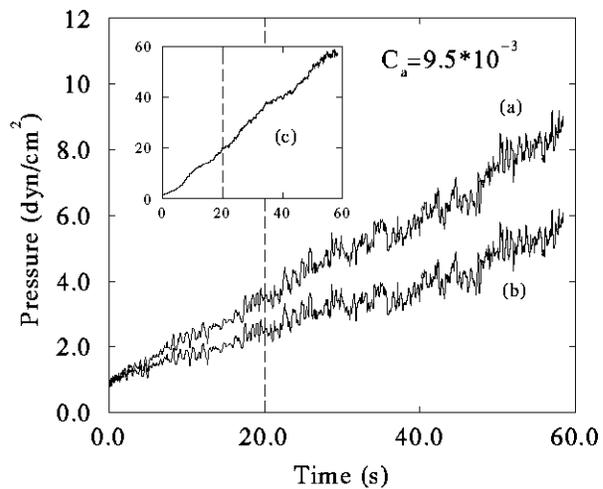
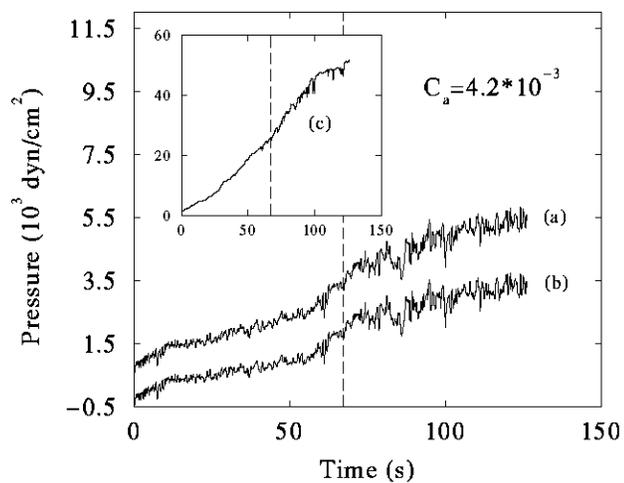
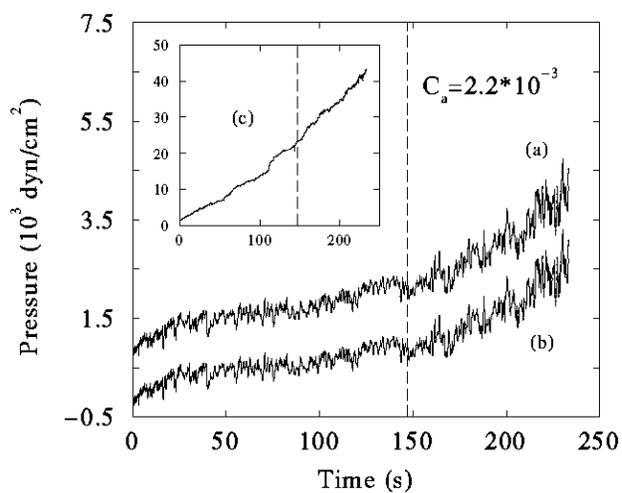
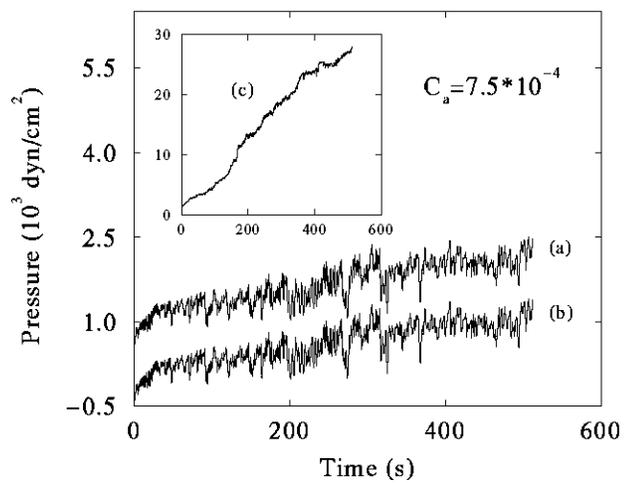
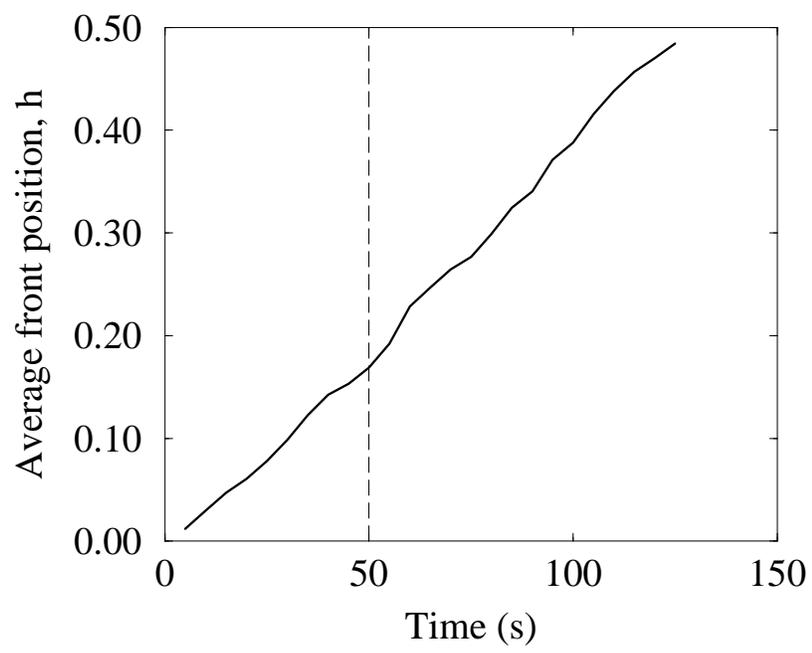
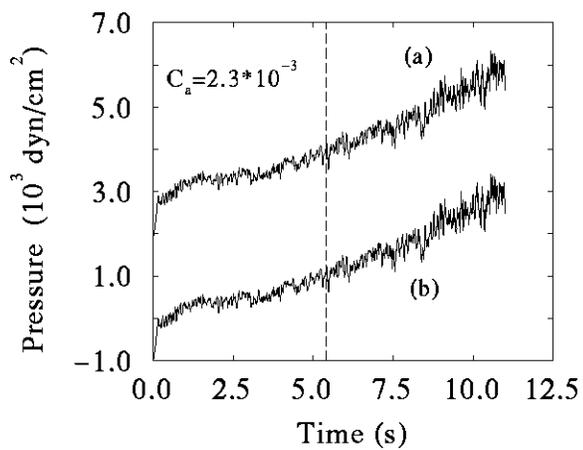
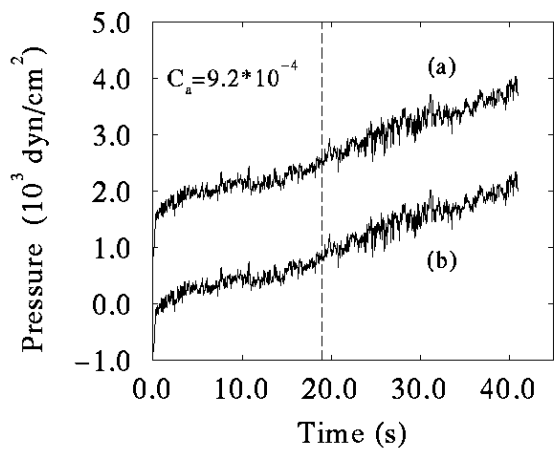
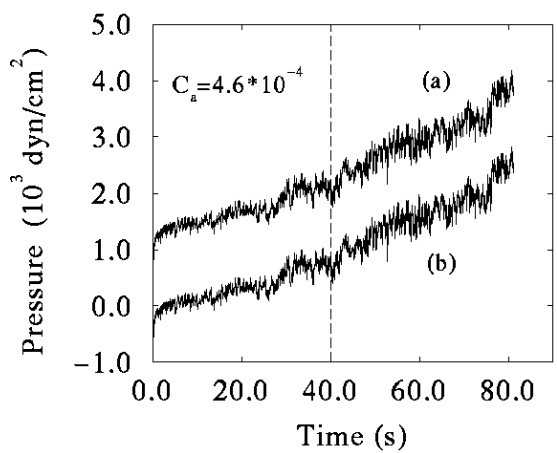
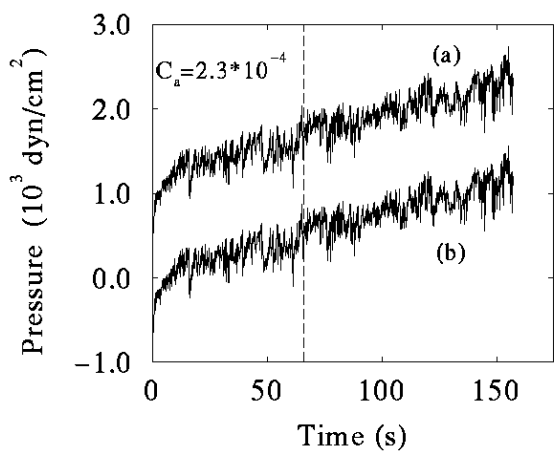
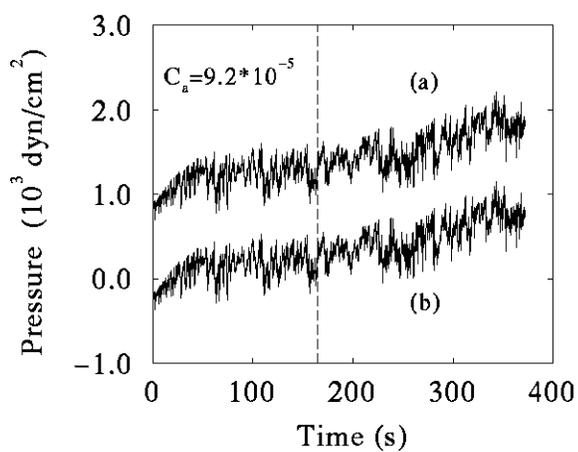
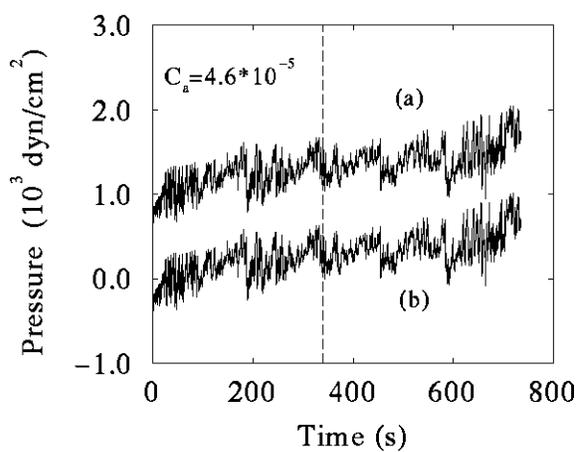
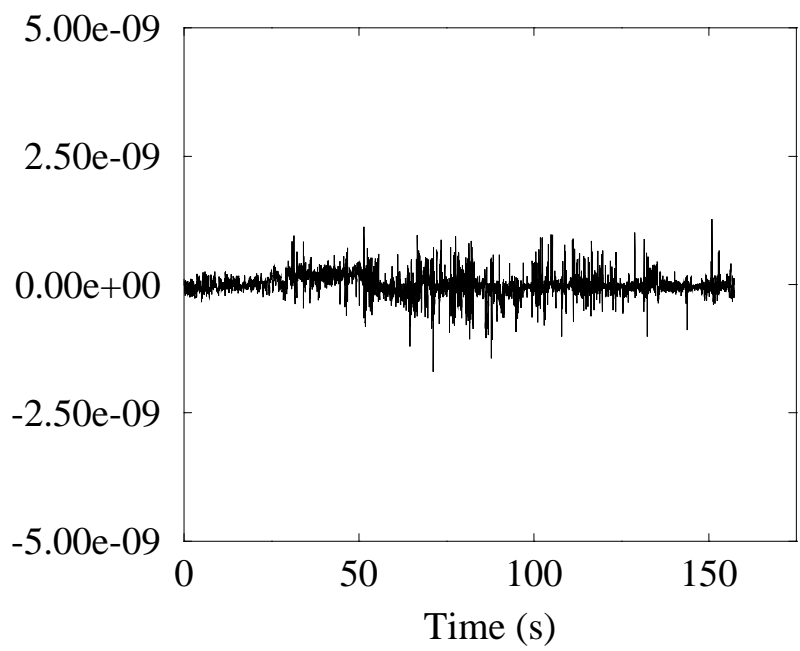
\end{document}